\documentclass[aps,prb,floatfix,twocolumn,superscriptaddress,a4paper]{revtex4}

\usepackage{mje,color}
\usepackage[abs]{overpic}
\usepackage[usenames,dvipsnames]{xcolor}

\newcommand{\NEW}[1]{#1}
\newcommand{\NEWM}[1]{#1}
\newcommand{\NEWN}[1]{#1}
\newcommand{\NEWO}[1]{#1}
\begin{document}

\title{Cool for Cats}

%% Notice placement of commas and superscripts and use of &
%% in the author list

\author{M.J. Everitt}
\email{m.j.everitt@physics.org}
\affiliation{Department of Physics, Loughborough University, Loughborough, Leics LE11 3TU, United Kingdom}
\author{T.P. Spiller}
\affiliation{Quantum Information Science, School of Physics and Astronomy, University of Leeds, Leeds LS2 9JT, United Kingdom}
\author{G.J. Milburn}
\affiliation{\NEW{Centre for Engineered Quantum Systems, School of Mathematics and Physics, The University of Queensland,} St Lucia, QLD 4072, Australia}
\author{R.D. Wilson}
\affiliation{Department of Physics, Loughborough University, Loughborough, Leics LE11 3TU, United Kingdom}
\author{A.M. Zagoskin}
\affiliation{Department of Physics, Loughborough University, Loughborough, Leics LE11 3TU, United Kingdom}

%\begin{document}

%\maketitle

%\begin{affiliations}
% \item 
% \item % \item 
%\end{affiliations}

\begin{abstract}

The iconic Schr\"odinger's cat state describes a system that may be in a superposition of two macroscopically distinct states, for example two clearly separated oscillator coherent states.  Quite apart from their role in understanding the quantum classical boundary,  such states have been suggested as offering a quantum advantage for quantum metrology, quantum communication and quantum computation. As is well known these applications have to face the difficulty that  the irreversible interaction with an environment causes the superposition to rapidly evolve to a mixture of the component states in the case that the environment is not monitored. Here we show that by engineering the interaction with the environment  \NEW{there exists a large class of systems that} can evolve irreversibly to a cat state.  To be precise we show that it is possible to engineer an irreversible process so that the steady state is close to a pure Schr\"odinger's cat state by using \NEWO{double well systems} and an environment comprising two-photon (or phonon) absorbers. We also show that it should be possible to prolong the lifetime of a Schr\"odinger's cat state exposed to the destructive effects of a conventional single-photon decohering environment. Our protocol should make it easier to prepare and maintain Schr\"odinger cat states which would be useful in applications of quantum metrology and information processing as well as being of interest to those probing the quantum to classical transition. 
\end{abstract}

\maketitle

The development of many quantum technologies depends on an ability to engineer strongly non classical states. Such states take the form of either highly entangled states of distinct degrees of freedom or a quantum coherent superposition of macroscopically distinct states in a single degree of freedom\cite{Sanders_review}, known as Schr\"odinger's cat states (after a well known thought experiment\cite{Schrodinger1935}). It is these cat states that we consider in this letter.  There has been great progress in the production of such states as well as experimentally reconstructing such states through a series of measurements in a process known of as quantum state tomography\cite{Haroche2008,Gao2010,Leibfried2005,Monroe1996,Noel1996,Ourjoumtsev2007}. 
These developments are of great importance as, in addition to their curious nature, Schr\"odinger cat states can be used as a resource for developing technologies such as quantum computing \cite{cat-comp1,cat-comp2},  quantum communication\cite{cat-comm1,cat-comm2} and quantum metrology~\cite{qmet1,qmet2,cat-metrology}. 
The main obstacle to deploying cat states in such applications is their fragility as they are destroyed by  noise in a process termed environmental decoherence. A careful consideration of optical cat states shows that this decoherence may be interpreted as due to Poisson distributed jumps between even and odd cat states whenever a single photon is lost\cite{Carmichael_stat,Carmichael_open,Vitali}. Their production and maintenance requires very precise quantum control as well as low dissipation. In this work we propose a  protocol \NEW{for \NEWO{double well} systems to create} Schr\"odinger cat states  that actually uses the non-controllable, non-unitary interaction of the system with a special kind of environment to create Schr\"odinger's cat states. To be specific, we have found that for a simple double-well system system interacting with an environment comprising a bath of two-photon absorbers, for certain initial states,  the system relaxed to a steady state which is close to a pure Schr\"odinger cat state. \NEW{Such an environment when paired with a parametric photon pump is known to exhibit many interesting effects in quantum optical systems, from cats to quantum statistics\cite{PhysRevA.49.2785,PhysRevA.48.1582}.} \NEWN{Two-photon absorption has also be suggested as a powerful resource for quantum computing application\cite{PhysRevA.70.062302}.} Two-photon decay preserves parity and enables the system to relax to  a steady state with same parity as the initial state. Our model is simpler than and  different from  other driven dissipative bistable systems (for example, the coherently driven optical cavity containing a Kerr medium\cite{Walls_Milb}, the driven Duffing mechanical resonator\cite{Katya}\NEWM{, tapered optical fibers\cite{PhysRevLett.105.173602}} and \NEW{photon pumps\cite{PhysRevA.49.2785,PhysRevA.48.1582}}),  as we do not include driving on either the cavity resonance or the coordinate degree of freedom. \NEW{Our proposal opens up new opportunities for exploring quantum phenomena from the micro to macroscopic level and in fields as diverse as quantum optics \cite{PhysRevLett.85.3365}, Boes-Einstein condensates \cite{Andrews1997}, quantum electronics \cite{Friedman:2000p1546} and nano-mechanics \cite{Badzey2005} \NEWN{(for which multi-phonon relaxation has already been proposed\cite{Voje:1302.1707})} or any other system in which it is possible to generate a \NEWN{double well} potential. }

\begin{figure}[!b]
\begin{center}
\includegraphics[width=0.45\textwidth]{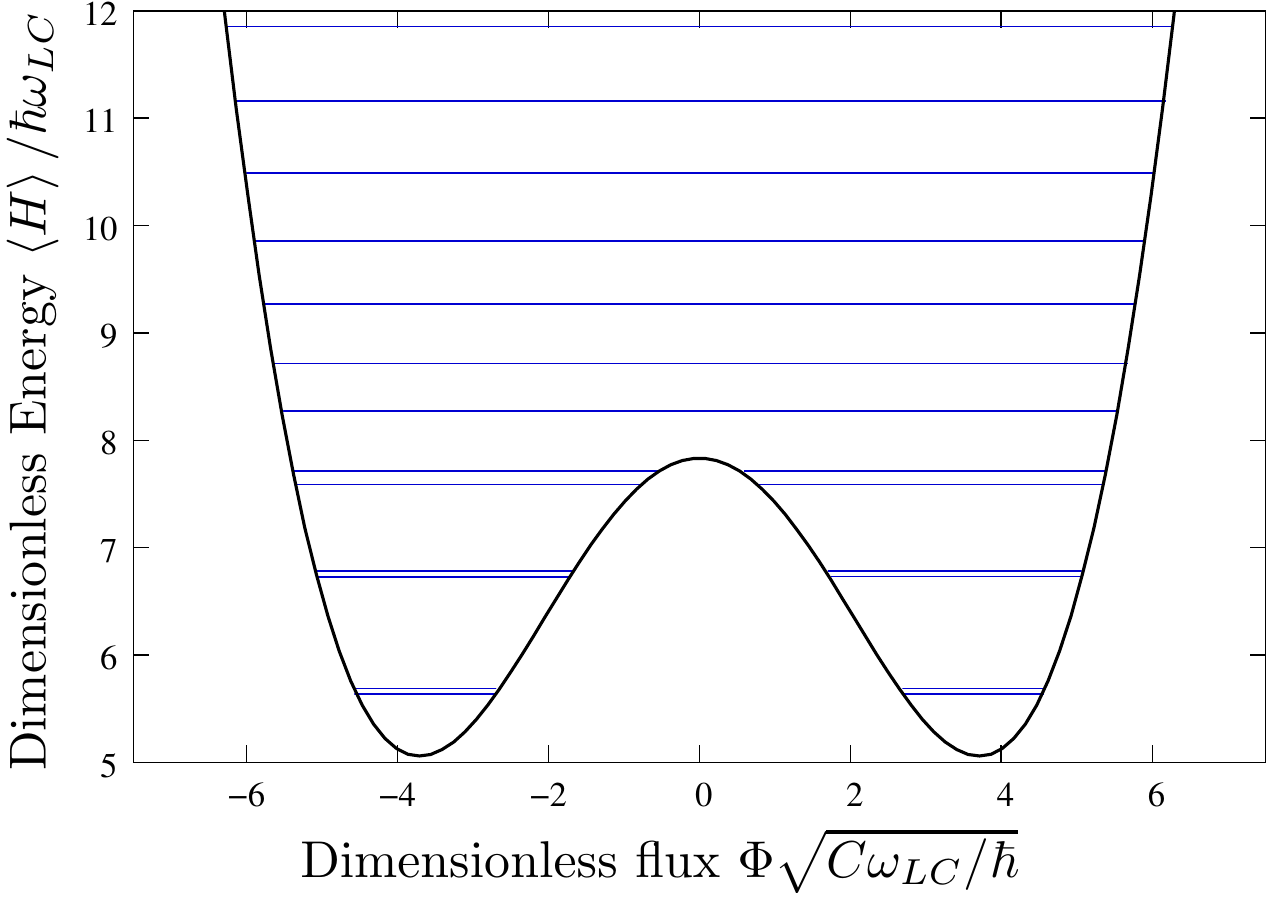}
\caption{\textbf{Stationary state energy levels:} The potential energy of the ring (black) as well as the energy of the rings stationary states (blue). Parameters used  here and throughout the paper are inductance $\Lambda=3\times 10^{-10}$H,  capacitance $C=5\times 10^{-15}$H, critical current of the weak link $I_{c}=2\mathrm{\mu A}$ and externally applied magnetic flux $\Phi_{x}=0.5\Phi_{0}$. Note that we have exaggerated the energy difference between the ground and first excited states as well as stationary states two and three in oder to make the different energies visible on this plot.}
\label{Fig:energyLevels} 
\end{center}
\end{figure}

For the results presented in this paper we have used  as an example system a superconducting quantum interference device (SQUID) ring. Our reason for choosing SQUIDs is that these devices are routinely fabricated and their theory is very well understood. We note that we have investigated a number of other systems (but do not include results here) and our analysis indicates that the key feature of the ring is that it can be made \NEWN{to form  a double well potential}. Moreover,
nonlinear systems derived from the Josepheson junction in circuit QED exhibit multi photon resonance when driven by an external field~\cite{2008NatPhysics.4.686} and thus we expect two-photon decay to be \NEWN{present} in such systems. The real difficulty is making it dominate over single photon effects. We will return to this later.
Beyond these considerations we believe there is nothing particularly special about 
\NEWN{the exact form of the potential needed to realise} our protocol.  \NEWN{Subject to being able to engineer an appropriate dissapative channel }\NEW{we therefore believe that the methodology that we propose for generating cat states will, as previously mentioned, find wide application.}  The potential energy of the SQUID comprising a thick superconducting ring enclosing a Josephson junction weak link takes the form of a harmonic oscillator perturbed by a cosine
$$
U(\Phi_{x})=\frac{(\Phi-\Phi_{x})^{2}}{2\Lambda}-\frac{\hbar I_{c}}{2e} \cos\left(2\pi \frac{\Phi}{\Phi_{0}} \right)
$$
where the coordinate $\Phi$ is the total magnetic flux  in the ring and $\Phi_{0}=h/2e$ is the superconducting flux quantum. We have chosen example circuit parameters that are in-line with modern fabrication techniques and suited to experimental realisations: $\Lambda=3\times 10^{-10}$H for the ring's inductance and $I_{c}=2\mathrm{\mu A}$ as the critical current of the weak link (although not in the above formula we also chose a capacitance $C=5\times 10^{-15}$F). We set the externally applied magnetic flux $\Phi_{x}=0.5\Phi_{0}$ so that the ring's potential forms a degenerate double well. It is also convenient to introduce the bosonic annihilation $a$, and creation $a^\dagger$ operators
$
\Phi =  \sqrt{\frac{\hbar}{2\NEWN{C}\omega_{LC}}}\left(a + a^\dagger\right)
$
where $\omega_{LC} = 1/\sqrt{\Lambda C}$.
%Although not shown in this work we have verified qualitative agreement with the key results in this paper for  example real circuit parameters from~\cite{Friedman:2000p1546}. Our choice of parameters was governed by optimising computational performance, the deeper wells of ~\cite{Friedman:2000p1546} lead to more demanding computations and provides no more insight.
In Fig.~\ref{Fig:energyLevels} we show the potential energy of the ring as well as the energy of the ring's stationary states. It is worth noting that the ground state and first excited state approximate, respectively, symmetric  and anti-symmetric superpositions of two coherent states centred at the bottom of each well. These two states have very nearly the same energy and the difference in their energy has been exaggerated in this plot (as have those for the second and third excited states).

We model the effect of the environment on the system using the master equation in the Lindblad form\NEWN{\cite{Viola199723}}
$$
\frac{\ud \rho}{\ud t}=-\frac{i}{\hbar}\COM{H}{\rho}+\frac{1}{2} \sum_{j}\left\{\COM{L_{j}}{\rho L_{j}^{\dag}}+\COM{L_{j} \rho}{L_{j}^{\dag}} \right\}
$$
where $\rho$ is the density matrix  describing the state of the system (initially $\rho=\OP{\psi(t=0)}{\psi(t=0)}$) and $H$ is the system's Hamiltonian. The non-unitary effect of the environment on the system is contained in the Lindbald operators $L_{j}$ with each describing a possible  environment. For example the usual Ohmic (i.e. analogous to friction proportional to velocity) bath, at zero temperature, would be described by a Lindblad proportional to the annihilation operator. For an undriven system the master equation has steady state solution that, in the presence of an  environment, is usually a density operator in a mixed state. In certain  circumstances, at zero temperature, these solutions may be  pure states such as the vacuum state of the harmonic oscillator. In these circumstances the solutions will not exhibit  features such as \NEWO{superpositions of macroscopically distinct states} and are relatively uninteresting. It is precisely this process where the environment essentially  removes the system's quantum coherence from \NEWO{de-localised}, or more generally non-Gaussian, states that is known of as environmental decoherence.  The density matrix for a decohered system without these quantum correlations represents a statistical mixture of possible states of the system and\NEWN{, for a single quantum object, } can be directly compared with classical probability density distributions\NEWN{\cite{PhysRevLett.80.4361}}. It should be noted however that there are driven dissipative systems, for example dispersive bistability, for which the steady state is a mixed state with a considerable amount of quantum coherence in the limit of large Kerr nonlinearity\cite{PhysRevLett.60.1836,Carmichael_stat,Carmichael_open}.

\begin{figure}[!b]
\begin{center}
\includegraphics[width=0.45\textwidth]{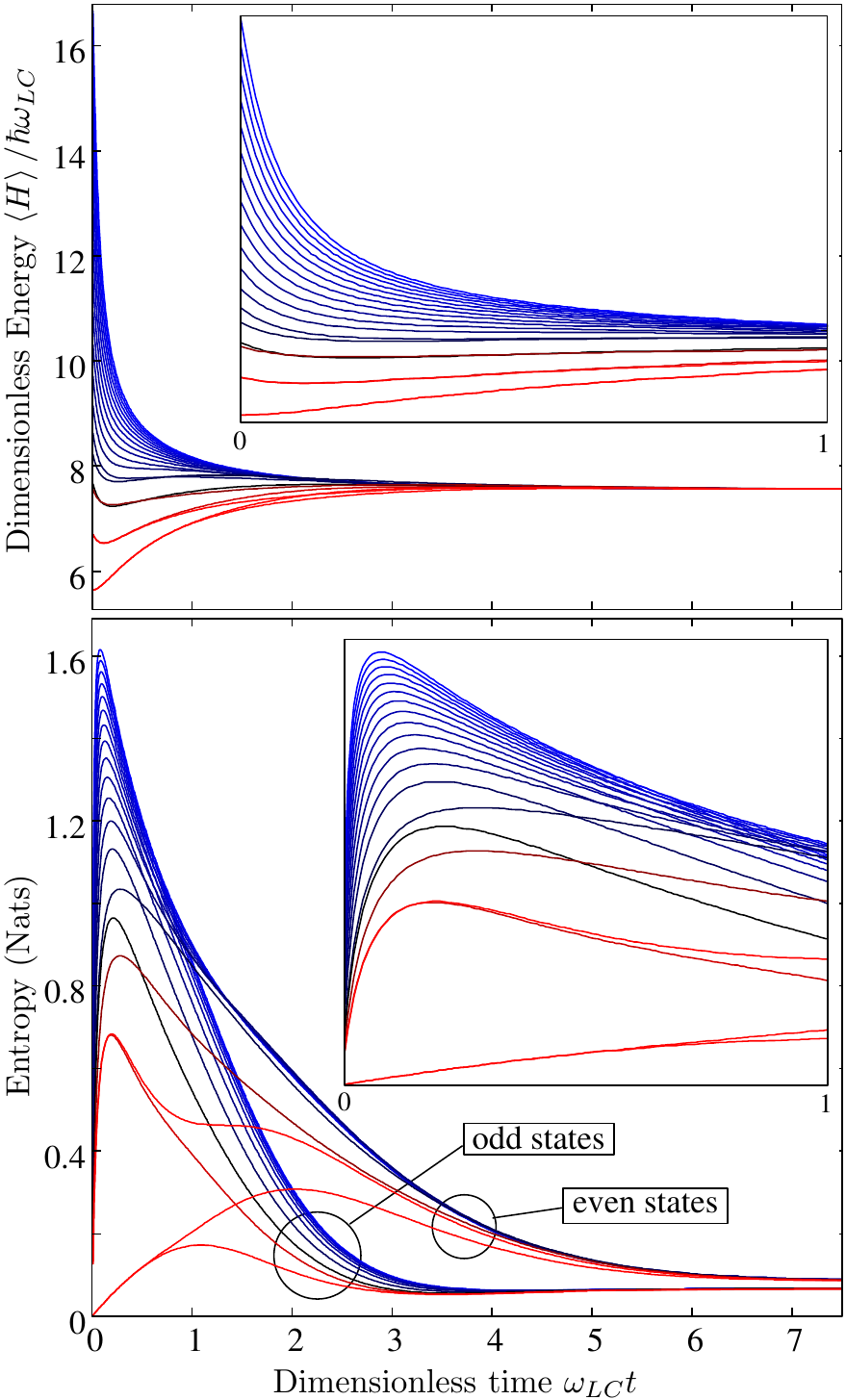}
\caption{\textbf{Effect of decoherence on energy and entropy} We show the dynamical evolution of the ring's energy and entropy using each of its first twenty stationary states as initial conditions. The dynamics have been found by solving the master equation for the ring in the presence of a bath of two-photon absorbers (with $L=\sqrt{0.2}a^{2}$). We have provided insets for increased resolution of the system's initial dynamics. The top plot shows the dynamics of the ring's total energy. As expected for an open quantum system of this kind the ring can be seen to decohere to one  energy, a little above that of the ground state. The bottom plot shows the dynamics of the von-Neuman entropy for the ring. In each case the initial entropy is zero as the system starts in a pure state. The entropy grows before dropping off to a low value indicating that the systems steady state solution is very nearly a pure state.}
\label{Fig:energyentropy}   
\end{center}
\end{figure}

We  found very different behaviour  if one chooses a different environment comprising two-photon absorbers, described by a Lindblad proportional to the square of the annihilation operator. In \Fig{Fig:energyentropy} 
we show the energy expectation values and von-Neumann entropy, $S=-\mathrm{Tr}[\rho \ln \rho]$ as functions of time for solutions of the master equation for the ring in the presence of such an environment.
 We used as initial conditions the first twenty energy eigenstates of the ring Hamiltonian. In these plots the energy behaves just as one would expect the energy of an undriven open quantum system to do -- it settles to a single value. When one inspects the dynamics of the entropy however the story is quite different. One \NEW{usually} expects the entropy to grow from zero to some asymptotic value as the system evolves into a mixed state. While we see that this is the initial behaviour the entropy does not monotonically increase, instead it decreases until the entropy is nearly negligible. It appears that the the system has to a significant extent recohered and the final density matrix is very nearly that of a pure state. While this is not the \NEW{usual} behaviour of an open quantum system it is in-line with our expectations of an environment that ``decoheres'' a system to a\NEWN{n almost pure state that is a very good approximation to a} Schr\"odinger cat state\NEW{\cite{PhysRevA.49.2785}}.

\begin{figure*}[!tb]
\begin{center}
\includegraphics[width=\textwidth]{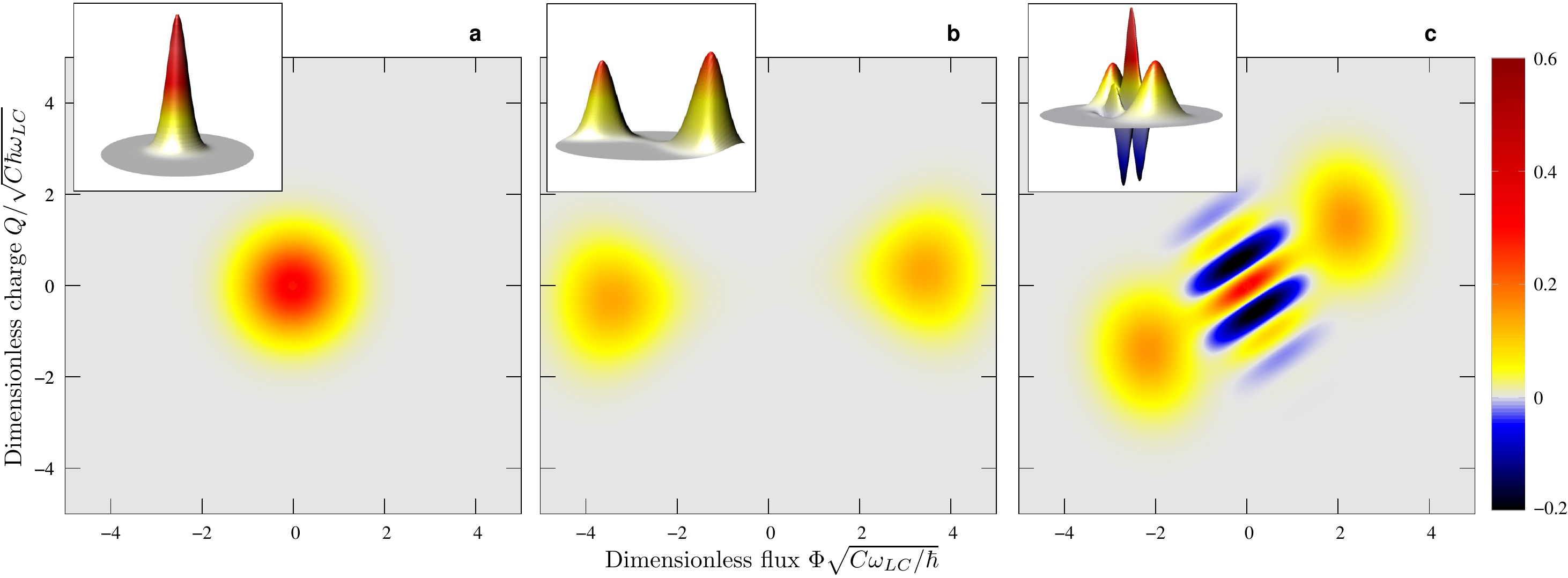}
\caption{\textbf{Cooling for a cat} In this figure we show, by making use of Wigner functions, the effect of two different environments on a ring prepared in a coherent state biased at zero flux. Each graph contains a top down view with a three dimensional  plot of the function as a not to fixed scale inset.  The graphs show \textbf{\textsf{a,}} the initial state which takes the form of a Gaussian bell. \textbf{\textsf{b,}} the steady state solution to the master equation under the influence of a conventional decohering environment comprising a  \NEWN{lossy} bath (with a Lindblad proportional to the annihilation operator $L=\sqrt{0.2}a$). The ring has decohered to two distinct macroscopic states we do not see the interference terms between them that are characteristic of a Schr\"odinger cat state. We have instead a statistical mixture, the usual and expected result~\cite{PhysRevA.69.043804}. \textbf{\textsf{c,}} the steady state solution to the master equation for the ring coupled to a bath of two-photon absorbers (with a Lindblad proportional to the square of annihilation operator $L=\sqrt{0.2}a^2$). In this case the ring has decohered to a superposition of two macroscopic states but now there are  interference terms between these states indicating quantum coherence - the signature of a Schr\"odinger cat state.}
\label{Fig:CoolToCat}   
\end{center}
\end{figure*}
In order to demonstrate that the system does indeed \NEWN{decay} to a Schr\"odinger cat state we will make use of the Wigner function. These pseudo probability density functions in phase space have been of great utility in demonstrating that some quantum states are Schr\"odinger cats\cite{Haroche2008}. The Wigner function is
$$
W(\Phi,Q)=\frac{1}{2\pi\hbar}\int \ME{\Phi+{\zeta}}{\rho}{\Phi-{\zeta}}\exp\left(-\frac{2iQ\zeta}{\hbar}\right) \ud \zeta
$$
where $Q$ is the \NEWN{charge} variable that is conjugate to the magnetic flux $\Phi$. In \Fig{Fig:CoolToCat} we show three Wigner functions.  
\Fig{Fig:CoolToCat}a shows the initial state and is a coherent state centred at the origin. This is clearly recognisable as the expected Gaussian bell shape associated with coherent states. We have solved the master equation for the ring in  a \NEWN{lossy} bath, with a Lindblad of $L=\sqrt{0.2}a$ and allowed the system to reach its steady state to obtain \Fig{Fig:CoolToCat}b. This is the  Wigner function of a statistical mixture of two macroscopically distinct states and is in-line with expectations of the effect of a decohering environment on such a device\cite{PhysRevA.69.043804}. In \Fig{Fig:CoolToCat}c we show the Wigner function that we obtain by solving the master equation, as for (b), but replacing the \NEWN{damping term} with a bath of two-photon absorbers, with $L=\sqrt{0.2}a^2$. We notice two things: firstly that the state has rotated which we believe to be a consequence of a squeezing action associated with the bath and secondly that there are interference terms between the distinct states of the system. These interference terms, indicating quantum coherence, confirm that this state state is indeed a \NEWN{very good approximation to a} Schr\"odinger cat. 
\begin{figure}[!b]
\begin{center}
\includegraphics[width=0.45\textwidth]{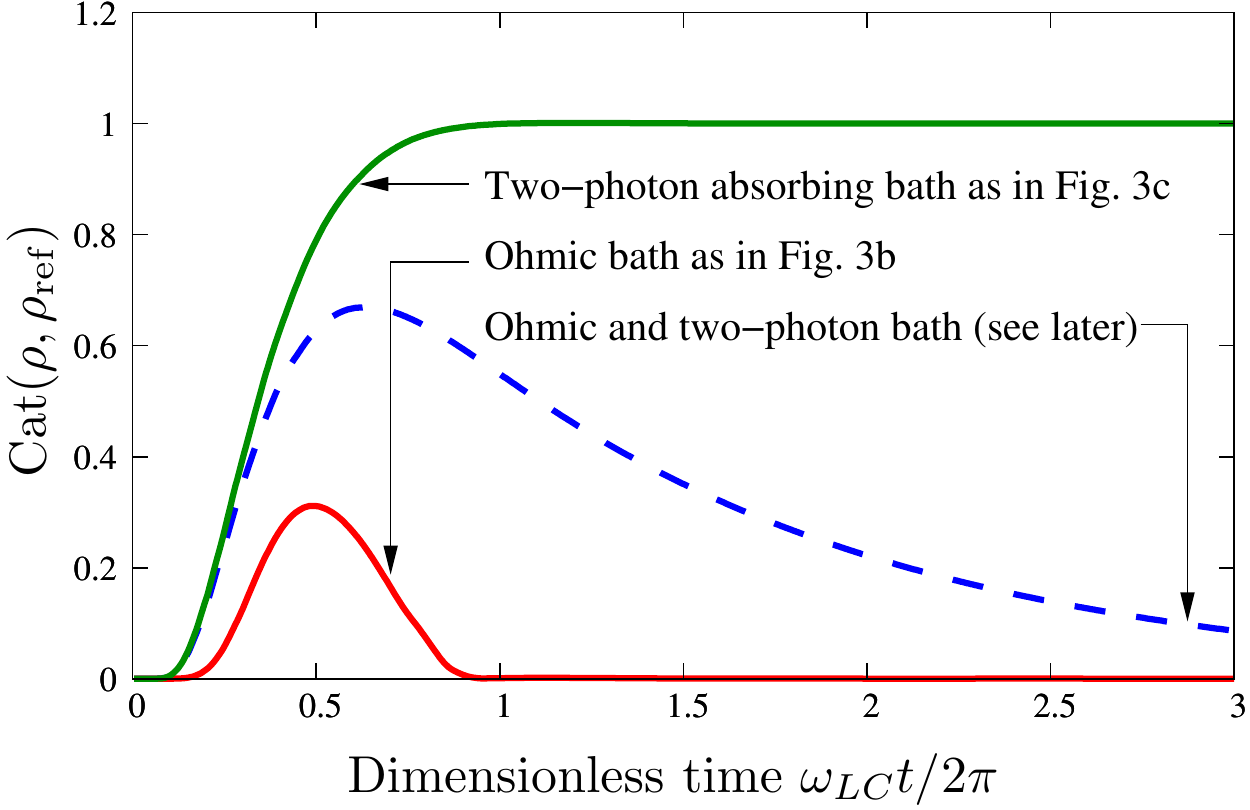}
\caption{\textbf{Relative cattiness} We show the cattiness measure $\mathrm{Cat(\rho,\rho_\mathrm{ref})}$ for the the dynamics leading to \Fig{Fig:CoolToCat}b in red and to \Fig{Fig:CoolToCat}c in green. Here we have used as a reference state $\rho_\mathrm{ref}$ the final cat state shown in \Fig{Fig:CoolToCat}c. For  reference later we have also included the dynamics of $\mathrm{Cat(\rho,\rho_\mathrm{ref})}$ for an environment of two-photon absorbers and damping.}
\label{Fig:Cat}   
\end{center}
\end{figure}
In oder to examine quantitatively the emergence of this cat from the initial coherent state we introduce, following~\cite{PhysRevA.62.054101,Biaynicki-Birula2002}, a measure  \NEWO{of how de-localised the system is in phase space} that is the integral of negative parts of the Wigner function
$$
	N(\rho)=\Half \int \left\{ |W(\Phi,Q)|- W(\Phi,Q) \right\} \, \ud \Phi \ud Q.
$$
In absolute terms this is a useful measure, but when we know (by inspecting the Wigner function) that the states we are examining are cat-like a more useful measure may well be a relative cattiness to some reference Schr\"odinger cat state. Hence we define:
$$
	\mathrm{Cat(\rho,\rho_\mathrm{ref})}=\frac{N(\rho)}{N(\rho_\mathrm{ref})}
$$
which quantifies the ratio of the \NEWO{de-localisation} of one cat state against a reference cat and enables us to quantify if one is more [$\mathrm{Cat(\rho,\rho_\mathrm{ref})}>1$], less [$\mathrm{Cat(\rho,\rho_\mathrm{ref})}<1$] or just as [$\mathrm{Cat(\rho,\rho_\mathrm{ref})}=1$] catty than the other. In \Fig{Fig:Cat} we show the dynamics of this quantity for comparison with the results presented in \Fig{Fig:CoolToCat} using as a reference state $\rho_\mathrm{ref}$ the final cat state shown in \Fig{Fig:CoolToCat}c. Here we can clearly see that the cattiness of the system subject to an environment of two-photon absorbers monotonically increases and asymptotically converges to a steady state. 

\begin{figure}[!tb]
\begin{center}
\includegraphics[width=0.45\textwidth]{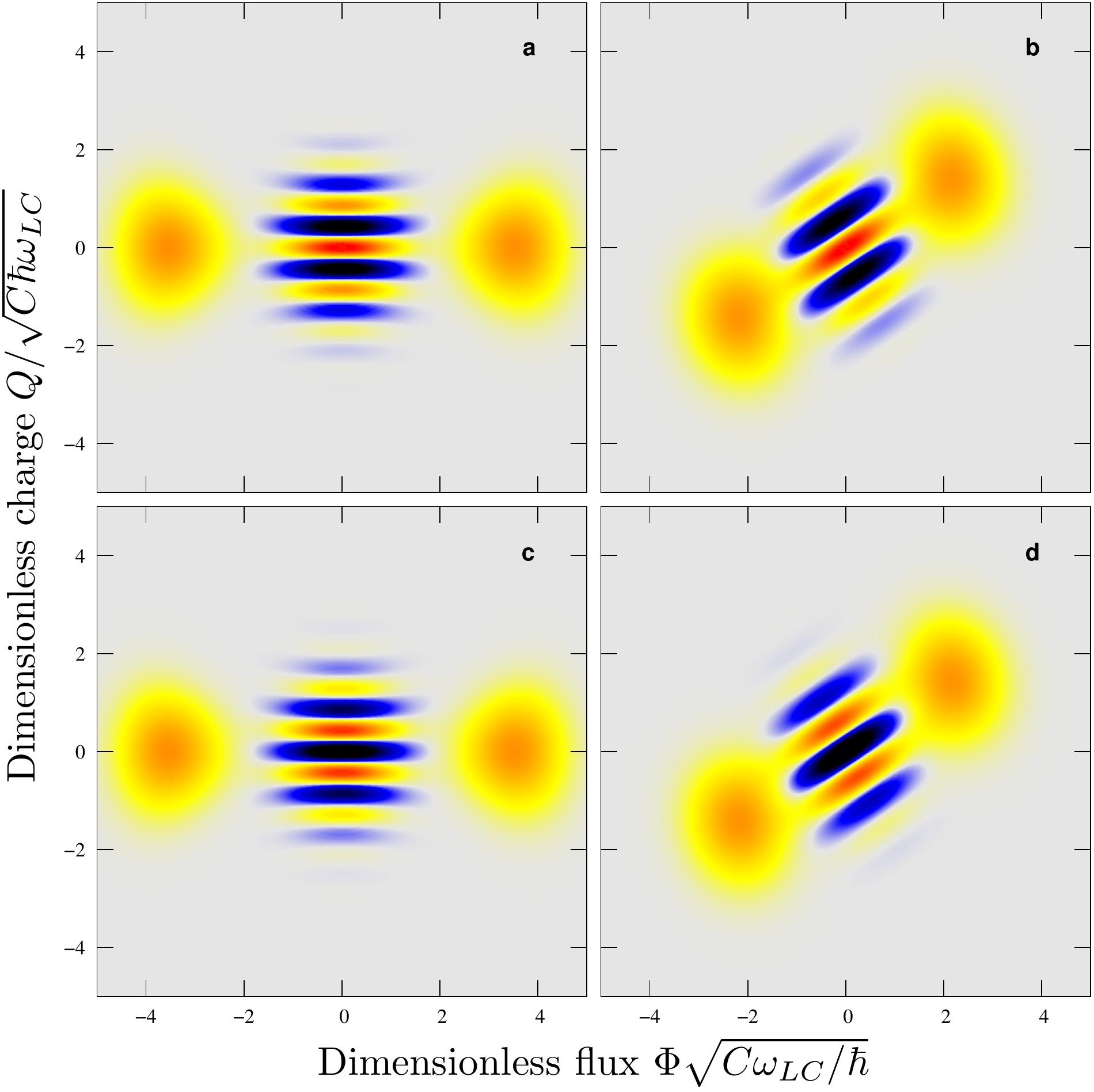}
\caption{\textbf{Preserving a cat} Here we look at the ring initially in either its ground or first excited stationary state. As can be seen from their Wigner functions, plots \textbf{\textsf{a,}} and \textbf{\textsf{c,}} respectively, these take the form of Schr\"odinger cat states. The ground state is, to good approximation, an even superposition of  two macroscopically distinct coherent states while the first excited state is an odd superposition. In terms of the Wigner functions this is reflected in the phase of the interference terms between the two Gauusian's of the cat. The effect of evolving the system in the presence of a bath of two-photon absorbers ($L=\sqrt{0.2}a^2$) is then shown with \textbf{\textsf{a}} evolving to \textbf{\textsf{b}} and \textbf{\textsf{c}} to \textbf{\textsf{d}}. We observe that the phase in the final cat reflects that of the initial cat and the system has not simply decohered to the same steady state.}
\label{Fig:PresCat}   
\end{center}
\end{figure}
It is interesting to consider what would happen to a ring that was initially in a Schr\"odinger cat state under the influence of a bath of two-photon absorbers. For systems with deep enough double well potentials such as the one considered here the ground and first excited energy eigenstates are both Schr\"odinger cats. The ground state is, to good approximation, an even superposition of  two macroscopically distinct coherent states while the first excited state is an odd superposition as can be seen from their Wigner functions in \Fig{Fig:PresCat}a and c respectively. The even and odd nature of these superpositions is reflected in the Wigner function by the phase of the interference terms between the two Gaussian's of the cat. It is known that such states would decohere under the environment of a  \NEWN{lossy} bath to a statistical mixture\cite{PhysRevA.69.043804}. The dynamics of the system coupled to an environment comprising a bath of two-photon absorbers are, once more, found by solving the master equation with an $L=\sqrt{0.2}a^{2}$, until an approximate steady state is reached.  The Wigner function of these states is then shown with \Fig{Fig:PresCat}a evolving to b and \Fig{Fig:PresCat}c to d. We observe that the phase in the final cat reflects that of the initial cat and the system has not simply decohered to the same steady state. The environment thus seems to preserve some of the symmetry of the initial state. We have checked the first twenty stationary states all of which decay to one of these cats or the other. Moreover, the pattern that was observed from the ground and first excited state persists and all even and odd states seem to evolve to cats of the same form as those shown \Fig{Fig:PresCat}b and \Fig{Fig:PresCat}d that are out of phase with each other.

\begin{figure*}[!tb]
\begin{center}
\includegraphics[width=\textwidth]{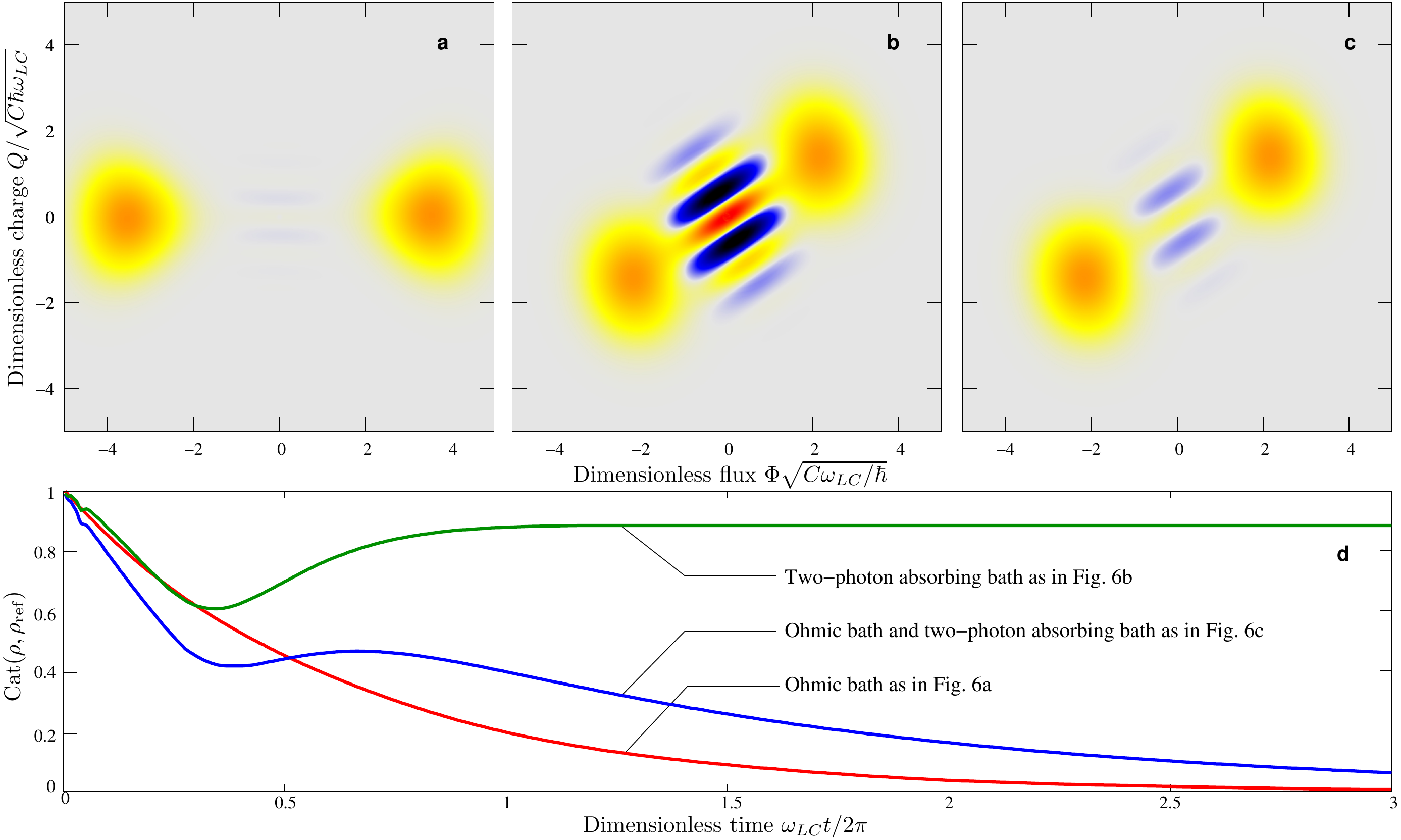}
\caption{\textbf{A stubborn cat: combatting the effect of other forms of decoherence} For each plot the system was initialised in its ground state of the ring as in \Fig{Fig:PresCat}a. In these plots we show \textbf{\textsf{a,}} the effect of a  \NEWN{lossy} bath on the state reducing a typical plot of a cat that has  just decohered to a statistical mixture - setting the time that we use to sample the other two plots of this figure. ($L=\sqrt{0.02}a$).  \textbf{\textsf{b,}} the effect of a two-photon absorbing bath showing decoherence to a Schr\"odinger cat state ($L=\sqrt{0.2}a^{2}$) \textbf{\textsf{c,}} the effect of both a  \NEWN{lossy} bath and a two-photon absorbing bath on the state. Notice that there are still signatures of a cat state unlike for the  \NEWN{lossy} bath alone -- the environment of two-photon absorbers seems to be prolonging the life of the cat  ($L_{1}=\sqrt{0.02}a$ and $L_{2}=\sqrt{0.2}a^{2}$) and \textbf{\textsf{d,}} we show the cattiness $\mathrm{Cat(\rho,\rho_\mathrm{ref})}$ for these three  environments as a function of time (we have used the initial stationary state as shown in \Fig{Fig:PresCat}a as the reference cat in this case). We see for the system's later evolution the environment of two-photon absorbers does indeed prolong the lifetime of the initial cat even in the presence of a  \NEWN{lossy} bath.}
\label{Fig:StubCat}   
\end{center}
\end{figure*}
Our protocol seems all very well and good but an environment of two-photon absorbers is very special. It would be hard to construct such an environment without having any other source of decoherence present. We therefore need to verify that the effects of a two-photon absorbing environment cannot be completely destroyed by the presence of a more traditional environment such as a \NEWN{lossy} bath. In \Fig{Fig:StubCat} we show the results of just such a check. For each plot the system's initial state was the ring's ground energy eigenstate as shown in \Fig{Fig:PresCat}a. In \Fig{Fig:StubCat}a we show the effect of a  \NEWN{lossy} bath. We solve the master equation with a Lindblad $L=\sqrt{0.02}a$ and allow the system to evolve until it has just decohered to a statistical mixture and we have plotted the Wigner function at this point in time. We use this run as a benchmark for computing the next two cases which show the Wigner function solutions of the master equation computed over the same interval. In \Fig{Fig:StubCat}b we show the effect of a two-photon absorbing bath once more ``decohering'' to a Schr\"odinger cat state ($L=\sqrt{0.2}a^{2}$). And in \Fig{Fig:StubCat}c we apply both the  \NEWN{lossy} bath of in \Fig{Fig:StubCat}a and a two-photon absorbing environment of in \Fig{Fig:StubCat}b to  the ring ($L_{1}=\sqrt{0.02}a$ and $L_{2}=\sqrt{0.2}a^{2}$).  We see that in this figure there remain residual Schr\"odinger cat state features in the Wigner function. Hence, it seems that not only does a  bath of two-photon absorbers create Schr\"odinger cat states, it also enables Schr\"odinger cat states to be more resilient to other forms of decoherence. In other words the presence of an environment of two-photon absorbers seems to be prolonging the life of a damped cat.
In \Fig{Fig:StubCat}d we quantify the cattiness using $\mathrm{Cat(\rho,\rho_\mathrm{ref})}$ using the initial stationary state as shown in \Fig{Fig:PresCat}a as the reference cat. For the three  environments considered here we find that for the system's later evolution the environment of two-photon absorbers does indeed prolong the lifetime of the initial cat even in the presence of a  \NEWN{lossy} bath.
 We note that we obtain an almost identical set of results if we start the system off in a coherent state centred at the origin (as in \Fig{Fig:CoolToCat}a). We chose to use the ring's ground state as, in our view, we obtained a more instructive plot of the states cattiness from the systems dynamics. For a direct comparison of the dynamics of $\mathrm{Cat(\rho,\rho_\mathrm{ref})}$ for these two initial conditions we now note that the dashed line shown in  \Fig{Fig:Cat} was found for a  \NEWN{lossy} bath and a two-photon absorbing environment  with $L_{1}=\sqrt{0.02}a$ and $L_{2}=\sqrt{0.2}a^{2}$. The green and blue lines of \Fig{Fig:Cat} and  \Fig{Fig:StubCat}d are directly comparable. The idea that the presence of a two-photon absorbing environment can be used to extend the lifetime (and also generate) Schr\"odinger cat states holds equally well for two very different initial conditions.
 
\NEWM{
In order to make our above discussion a reality we need to engineer a dissipative quantum channel that acts as a two-photon absorber. Here we suggest a concrete realisation that, whilst not perfect, still retains the key feature of environmentally induced ``decoherence'' to a Schr\"odinger cat state. Our proposal makes use of non-linearly coupled electromagnetic fields and SQUIDs. Such quantum electrodynamic circuits have already been investigated in the context of weak non-demolition measurement\cite{PhysRevB.82.014512,PhysRevB.82.220505}. \NEWN{One example comprises two microwave superconducting resonators coupled via a SQUID which in addition to a cross Kerr effect also manifests two photon conversion terms if the cavities are resonant}\cite{PhysRevB.82.014512}. Such systems can be quantised\cite{PhysRevB.74.224506,PhysRevB.63.144530,PhysRevB.64.184517,PhysRevB.72.014508} and with a suitable arrangement and choice of circuit parameters  can be reduced\cite{PhysRevA.48.2494,PhysRevA.48.2494,PhysRevA.70.052105} to the form of a double well system subject to a two-photon absorbing environment (see supplementary material for details). Unavoidably, this process also brings with it an additional dephasing term, that adds to the master equation another Lindblad proportional to $a^{\dag}a$. Nevertheless, we can report that whilst the dephasing term smears out the Gaussian peaks in the cat the interference terms in the Wigner function representing quantum coherence between the cat states remains strong. \NEWN{The fact that this dephasing term preserves parity is once more the key factor in ensuring}  the steady state of our engineered dissipative channel is still a Schr\"odinger cat state.  Our proposal could lead to an initial realisation of a two-photon absorbing environment and concomitant interesting effects. The engineering of improved dissipative channels, without additional and unwanted decoherence effects, remains an open and interesting problem.
}

\NEWN{There are two phenomena that embody quantum mechanics, namely entanglement and  the Schr\"odinger's cat thought experiment\cite{Schrodinger1935}. The latter was p}roposed to highlight the difficulties we have connecting quantum mechanics with everyday experience it neatly demonstrates the problems of understanding the emergence of the classical world from quantum theory and the measurement of quantum systems. Schr\"odinger's cat has become the icon of the subject and evolved to have a well defined meaning.  It is an accepted explanation within the \NEWN {popular} literature that the reason the original thought experiment does not translate into reality (if conducted with a real cat in a box etc.) is that the environment to the radiation source (which included the cat itself) deletes the quantumness connecting the two states in a process known as decorehence. As such
environmental decoherence is something that many deem to be a crucial element in the the quantum to classical transition~\cite{Bell,RevModPhys.76.1267,MJENJPQCT,PhysRevA.79.032328,PhysRevLett.80.4361}. We have shown that some environments may have a dramatically different effect on \NEWN{double well systems} producing very quantum states as a result of ``decoherence''. It may well be that \NEWN{system and} environment such as the one we have \NEW{used} here could play an interesting role in \NEWN{quantum mechanically enhanced metrology} probing foundational aspects of quantum mechanics \NEWN{associated with realising macroscopic quantum phenomena and the quantum to classical transition}. Furthermore, it is known that open quantum systems can be used to model the measurement process. Although it is beyond the scope of the current paper, we conjecture that it may well be possible to make use of an environment to measure a system into a Schr\"odinger cat state.  

%\bibliography{ref}

\begin{itemize}
 \item MJE, RDW, and AMZ thank the Templeton Foundation for their generous support. \NEW{GJM acknowledges the support of the Australian Research Council Centre of Excellence for Engineered Quantum Systems grant CE110001013. We would like to thank Peter Knight for interesting and informative discussions.} MJE would like to thank Andrew Archer, Gerry Swallowe and Richard Giles   for their  help with the preparation of our manuscript. 
 \item The authors declare that they have no competing financial interests.
 \item Correspondence and requests for materials should be addressed to M.J. Everitt~(email: m.j.everitt@physics.org).
 \item Authors' contributions: TS and MJE formulated the original problem and choice of the system (choice of system was independently corroborated by AZ);
GJM proposed the form of the environment \NEWM{and its possible realisation};
MJE directed and RDW performed numerical calculations and analysis of results;
All authors contributed to the technical discussion;
MJE and RDW prepared the manuscript with help from all other authors.
\end{itemize}
\bibliography{ref}

\begin{thebibliography}{10}
\expandafter\ifx\csname url\endcsname\relax
  \def\url#1{\texttt{#1}}\fi
\expandafter\ifx\csname urlprefix\endcsname\relax\def\urlprefix{URL }\fi
\providecommand{\bibinfo}[2]{#2}
\providecommand{\eprint}[2][]{\url{#2}}

\bibitem{Sanders_review}
\bibinfo{author}{Sanders, B.~C.}
\newblock \bibinfo{title}{Review of entangled coherent states}.
\newblock \emph{\bibinfo{journal}{Journal of Physics A: Mathematical and
  Theoretical}} \textbf{\bibinfo{volume}{45}}, \bibinfo{pages}{244002}
  (\bibinfo{year}{2012}).
\newblock \urlprefix\url{http://stacks.iop.org/1751-8121/45/i=24/a=244002}.

\bibitem{Schrodinger1935}
\bibinfo{author}{Schr\"odinger, E.}
\newblock \bibinfo{title}{Die gegenw\"artige situation in der quantenmechanik}.
\newblock \emph{\bibinfo{journal}{Naturwissenschaften}}
  \textbf{\bibinfo{volume}{23}}, \bibinfo{pages}{807--812}
  (\bibinfo{year}{1935}).
\newblock \urlprefix\url{http://dx.doi.org/10.1007/BF01491891}.

\bibitem{Haroche2008}
\bibinfo{author}{Del{\'e}glise, S.} \emph{et~al.}
\newblock \bibinfo{title}{Reconstruction of non-classical cavity field states
  with snapshots of their decoherence}.
\newblock \emph{\bibinfo{journal}{Nature}} \textbf{\bibinfo{volume}{455}}
  (\bibinfo{year}{2008}).

\bibitem{Gao2010}
\bibinfo{author}{Gao, W.-B.} \emph{et~al.}
\newblock \bibinfo{title}{{Experimental demonstration of a hyper-entangled
  ten-qubit Schr\"{o}dinger cat state}}.
\newblock \emph{\bibinfo{journal}{Nature Physics}}
  \textbf{\bibinfo{volume}{6}}, \bibinfo{pages}{331--335}
  (\bibinfo{year}{2010}).
\newblock \urlprefix\url{http://dx.doi.org/10.1038/nphys1603}.

\bibitem{Leibfried2005}
\bibinfo{author}{Leibfried, D.} \emph{et~al.}
\newblock \bibinfo{title}{{Creation of a six-atom 'Schr\"{o}dinger cat'
  state.}}
\newblock \emph{\bibinfo{journal}{Nature}} \textbf{\bibinfo{volume}{438}},
  \bibinfo{pages}{639--42} (\bibinfo{year}{2005}).
\newblock \urlprefix\url{http://dx.doi.org/10.1038/nature04251}.

\bibitem{Monroe1996}
\bibinfo{author}{Monroe, C.}, \bibinfo{author}{Meekhof, D.~M.},
  \bibinfo{author}{King, B.~E.} \& \bibinfo{author}{Wineland, D.~J.}
\newblock \bibinfo{title}{{A ``Schr\"odinger Cat'' Superposition State of an
  Atom}}.
\newblock \emph{\bibinfo{journal}{Science}} \textbf{\bibinfo{volume}{272}},
  \bibinfo{pages}{1131--1136} (\bibinfo{year}{1996}).
\newblock
  \urlprefix\url{http://www.sciencemag.org/content/272/5265/1131.abstract}.

\bibitem{Noel1996}
\bibinfo{author}{Noel, M.} \& \bibinfo{author}{Stroud, J.}
\newblock \bibinfo{title}{{Excitation of an Atomic Electron to a Coherent
  Superposition of Macroscopically Distinct States}}.
\newblock \emph{\bibinfo{journal}{Physical Review Letters}}
  \textbf{\bibinfo{volume}{77}}, \bibinfo{pages}{1913--1916}
  (\bibinfo{year}{1996}).
\newblock \urlprefix\url{http://link.aps.org/doi/10.1103/PhysRevLett.77.1913}.

\bibitem{Ourjoumtsev2007}
\bibinfo{author}{Ourjoumtsev, A.}, \bibinfo{author}{Jeong, H.},
  \bibinfo{author}{Tualle-Brouri, R.} \& \bibinfo{author}{Grangier, P.}
\newblock \bibinfo{title}{{Generation of optical 'Schr\"{o}dinger cats' from
  photon number states.}}
\newblock \emph{\bibinfo{journal}{Nature}} \textbf{\bibinfo{volume}{448}},
  \bibinfo{pages}{784--6} (\bibinfo{year}{2007}).
\newblock \urlprefix\url{http://dx.doi.org/10.1038/nature06054}.

\bibitem{cat-comp1}
\bibinfo{author}{Ralph, T.~C.}, \bibinfo{author}{Gilchrist, A.},
  \bibinfo{author}{Milburn, G.~J.}, \bibinfo{author}{Munro, W.~J.} \&
  \bibinfo{author}{Glancy, S.}
\newblock \bibinfo{title}{Quantum computation with optical coherent states}.
\newblock \emph{\bibinfo{journal}{Phys. Rev. A}} \textbf{\bibinfo{volume}{68}},
  \bibinfo{pages}{042319} (\bibinfo{year}{2003}).
\newblock \urlprefix\url{http://link.aps.org/doi/10.1103/PhysRevA.68.042319}.

\bibitem{cat-comp2}
\bibinfo{author}{Gilchrist, A.} \emph{et~al.}
\newblock \bibinfo{title}{Schrdinger cats and their power for quantum
  information processing}.
\newblock \emph{\bibinfo{journal}{Journal of Optics B: Quantum and
  Semiclassical Optics}} \textbf{\bibinfo{volume}{6}}, \bibinfo{pages}{S828}
  (\bibinfo{year}{2004}).
\newblock \urlprefix\url{http://stacks.iop.org/1464-4266/6/i=8/a=032}.

\bibitem{cat-comm1}
\bibinfo{author}{van Enk, S.~J.} \& \bibinfo{author}{Hirota, O.}
\newblock \bibinfo{title}{Entangled coherent states: Teleportation and
  decoherence}.
\newblock \emph{\bibinfo{journal}{Phys. Rev. A}} \textbf{\bibinfo{volume}{64}},
  \bibinfo{pages}{022313} (\bibinfo{year}{2001}).
\newblock \urlprefix\url{http://link.aps.org/doi/10.1103/PhysRevA.64.022313}.

\bibitem{cat-comm2}
\bibinfo{author}{Jeong, H.}, \bibinfo{author}{Kim, M.~S.} \&
  \bibinfo{author}{Lee, J.}
\newblock \bibinfo{title}{Quantum-information processing for a coherent
  superposition state via a mixed entangled coherent channel}.
\newblock \emph{\bibinfo{journal}{Phys. Rev. A}} \textbf{\bibinfo{volume}{64}},
  \bibinfo{pages}{052308} (\bibinfo{year}{2001}).
\newblock \urlprefix\url{http://link.aps.org/doi/10.1103/PhysRevA.64.052308}.

\bibitem{qmet1}
\bibinfo{author}{Blatt, R.} \& \bibinfo{author}{Wineland, D.}
\newblock \bibinfo{title}{Entangled states of trapped atomic ions}.
\newblock \emph{\bibinfo{journal}{Nature}} \textbf{\bibinfo{volume}{453}},
  \bibinfo{pages}{1008--1015} (\bibinfo{year}{2008}).

\bibitem{qmet2}
\bibinfo{author}{Giovannetti, V.}, \bibinfo{author}{Lloyd, S.} \&
  \bibinfo{author}{Maccone, L.}
\newblock \bibinfo{title}{Advances in quantum metrology}.
\newblock \emph{\bibinfo{journal}{Nat Photon}} \textbf{\bibinfo{volume}{23}},
  \bibinfo{pages}{222--229} (\bibinfo{year}{2011}).

\bibitem{cat-metrology}
\bibinfo{author}{Munro, W.~J.}, \bibinfo{author}{Nemoto, K.},
  \bibinfo{author}{Milburn, G.~J.} \& \bibinfo{author}{Braunstein, S.~L.}
\newblock \bibinfo{title}{Weak-force detection with superposed coherent
  states}.
\newblock \emph{\bibinfo{journal}{Phys. Rev. A}} \textbf{\bibinfo{volume}{66}},
  \bibinfo{pages}{023819} (\bibinfo{year}{2002}).
\newblock \urlprefix\url{http://link.aps.org/doi/10.1103/PhysRevA.66.023819}.

\bibitem{Carmichael_stat}
\bibinfo{author}{Carmichael, H.}
\newblock \emph{\bibinfo{title}{Statistical Methods in Quantum Optics 1: Master
  Equations and Fokker-Planck Equations}}.
\newblock Statistical Methods in Quantum Optics (\bibinfo{publisher}{Springer},
  \bibinfo{year}{2003}).
\newblock \urlprefix\url{http://books.google.co.uk/books?id=ocgRgM-yJacC}.

\bibitem{Carmichael_open}
\bibinfo{author}{Carmichael, H.}
\newblock \emph{\bibinfo{title}{An open systems approach to quantum optics:
  lectures presented at the Universit{\'e} Libre de Bruxelles, October 28 -
  November 4, 1991}}.
\newblock Lecture Notes in Physics Series
  (\bibinfo{publisher}{Springer-Verlag}, \bibinfo{year}{1993}).
\newblock \urlprefix\url{http://books.google.co.uk/books?id=El5gxgxWhpgC}.

\bibitem{Vitali}
\bibinfo{author}{Vitali, D.}, \bibinfo{author}{Tombesi, P.} \&
  \bibinfo{author}{Milburn, G.~J.}
\newblock \bibinfo{title}{Controlling the decoherence of a ``meter'' via
  stroboscopic feedback}.
\newblock \emph{\bibinfo{journal}{Phys. Rev. Lett.}}
  \textbf{\bibinfo{volume}{79}}, \bibinfo{pages}{2442--2445}
  (\bibinfo{year}{1997}).
\newblock \urlprefix\url{http://link.aps.org/doi/10.1103/PhysRevLett.79.2442}.

\bibitem{PhysRevA.49.2785}
\bibinfo{author}{Gilles, L.}, \bibinfo{author}{Garraway, B.~M.} \&
  \bibinfo{author}{Knight, P.~L.}
\newblock \bibinfo{title}{Generation of nonclassical light by dissipative
  two-photon processes}.
\newblock \emph{\bibinfo{journal}{Phys. Rev. A}} \textbf{\bibinfo{volume}{49}},
  \bibinfo{pages}{2785--2799} (\bibinfo{year}{1994}).
\newblock \urlprefix\url{http://link.aps.org/doi/10.1103/PhysRevA.49.2785}.

\bibitem{PhysRevA.48.1582}
\bibinfo{author}{Gilles, L.} \& \bibinfo{author}{Knight, P.~L.}
\newblock \bibinfo{title}{Two-photon absorption and nonclassical states of
  light}.
\newblock \emph{\bibinfo{journal}{Phys. Rev. A}} \textbf{\bibinfo{volume}{48}},
  \bibinfo{pages}{1582--1593} (\bibinfo{year}{1993}).
\newblock \urlprefix\url{http://link.aps.org/doi/10.1103/PhysRevA.48.1582}.

\bibitem{PhysRevA.70.062302}
\bibinfo{author}{Franson, J.~D.}, \bibinfo{author}{Jacobs, B.~C.} \&
  \bibinfo{author}{Pittman, T.~B.}
\newblock \bibinfo{title}{Quantum computing using single photons and the zeno
  effect}.
\newblock \emph{\bibinfo{journal}{Phys. Rev. A}} \textbf{\bibinfo{volume}{70}},
  \bibinfo{pages}{062302} (\bibinfo{year}{2004}).
\newblock \urlprefix\url{http://link.aps.org/doi/10.1103/PhysRevA.70.062302}.

\bibitem{Walls_Milb}
\bibinfo{author}{Walls, D.} \& \bibinfo{author}{Milburn, G.}
\newblock \emph{\bibinfo{title}{Quantum Optics}}.
\newblock SpringerLink: Springer e-Books (\bibinfo{publisher}{Springer},
  \bibinfo{year}{2008}).
\newblock \urlprefix\url{http://books.google.co.uk/books?id=LiWsc3Nlf0kC}.

\bibitem{Katya}
\bibinfo{author}{Babourina-Brooks, E.}, \bibinfo{author}{Doherty, A.} \&
  \bibinfo{author}{Milburn, G.~J.}
\newblock \bibinfo{title}{Quantum noise in a nanomechanical duffing resonator}.
\newblock \emph{\bibinfo{journal}{New Journal of Physics}}
  \textbf{\bibinfo{volume}{10}}, \bibinfo{pages}{105020}
  (\bibinfo{year}{2008}).
\newblock \urlprefix\url{http://stacks.iop.org/1367-2630/10/i=10/a=105020}.

\bibitem{PhysRevLett.105.173602}
\bibinfo{author}{Hendrickson, S.~M.}, \bibinfo{author}{Lai, M.~M.},
  \bibinfo{author}{Pittman, T.~B.} \& \bibinfo{author}{Franson, J.~D.}
\newblock \bibinfo{title}{Observation of two-photon absorption at low power
  levels using tapered optical fibers in rubidium vapor}.
\newblock \emph{\bibinfo{journal}{Phys. Rev. Lett.}}
  \textbf{\bibinfo{volume}{105}}, \bibinfo{pages}{173602}
  (\bibinfo{year}{2010}).
\newblock
  \urlprefix\url{http://link.aps.org/doi/10.1103/PhysRevLett.105.173602}.

\bibitem{PhysRevLett.85.3365}
\bibinfo{author}{Haycock, D.~L.}, \bibinfo{author}{Alsing, P.~M.},
  \bibinfo{author}{Deutsch, I.~H.}, \bibinfo{author}{Grondalski, J.} \&
  \bibinfo{author}{Jessen, P.~S.}
\newblock \bibinfo{title}{Mesoscopic quantum coherence in an optical lattice}.
\newblock \emph{\bibinfo{journal}{Phys. Rev. Lett.}}
  \textbf{\bibinfo{volume}{85}}, \bibinfo{pages}{3365--3368}
  (\bibinfo{year}{2000}).
\newblock \urlprefix\url{http://link.aps.org/doi/10.1103/PhysRevLett.85.3365}.

\bibitem{Andrews1997}
\bibinfo{author}{Andrews, M.~R.}
\newblock \bibinfo{title}{{Observation of Interference Between Two Bose
  Condensates}}.
\newblock \emph{\bibinfo{journal}{Science}} \textbf{\bibinfo{volume}{275}},
  \bibinfo{pages}{637--641} (\bibinfo{year}{1997}).
\newblock
  \urlprefix\url{http://www.sciencemag.org/content/275/5300/637.abstract}.

\bibitem{Friedman:2000p1546}
\bibinfo{author}{Friedman, J.~R.}, \bibinfo{author}{Patel, V.},
  \bibinfo{author}{Chen, W.}, \bibinfo{author}{Tolpygo, S.~K.} \&
  \bibinfo{author}{Lukens, J.~E.}
\newblock \bibinfo{title}{Quantum superposition of distinct macroscopic
  states}.
\newblock \emph{\bibinfo{journal}{Nature}} \textbf{\bibinfo{volume}{406}},
  \bibinfo{pages}{43} (\bibinfo{year}{2000}).

\bibitem{Badzey2005}
\bibinfo{author}{Badzey, R.~L.} \& \bibinfo{author}{Mohanty, P.}
\newblock \bibinfo{title}{{Coherent signal amplification in bistable
  nanomechanical oscillators by stochastic resonance.}}
\newblock \emph{\bibinfo{journal}{Nature}} \textbf{\bibinfo{volume}{437}},
  \bibinfo{pages}{995--8} (\bibinfo{year}{2005}).
\newblock
  \urlprefix\url{http://www.nature.com/nature/journal/v437/n7061/full/nature04124.html\#B9}.

\bibitem{Voje:1302.1707}
\bibinfo{author}{Voje, A.}, \bibinfo{author}{Croy, A.} \&
  \bibinfo{author}{Isacsson, A.}
\newblock \bibinfo{title}{{Multi-phonon relaxation and generation of quantum
  states in a nonlinear mechanical oscillator}}  (\bibinfo{year}{2013}).
\newblock \eprint{arXiv:1302.1707}.

\bibitem{2008NatPhysics.4.686}
\bibinfo{author}{{Deppe}, F.} \emph{et~al.}
\newblock \bibinfo{title}{{Two-photon probe of the Jaynes Cummings model and
  controlled symmetry breaking in circuit QED}}.
\newblock \emph{\bibinfo{journal}{Nature Physics}}
  \textbf{\bibinfo{volume}{4}}, \bibinfo{pages}{686} (\bibinfo{year}{2008}).
\newblock \eprint{0805.3294}.

\bibitem{Viola199723}
\bibinfo{author}{Viola, L.}, \bibinfo{author}{Onofrio, R.} \&
  \bibinfo{author}{Calarco, T.}
\newblock \bibinfo{title}{Macroscopic quantum damping in squid rings}.
\newblock \emph{\bibinfo{journal}{Physics Letters A}}
  \textbf{\bibinfo{volume}{229}}, \bibinfo{pages}{23 -- 31}
  (\bibinfo{year}{1997}).
\newblock
  \urlprefix\url{http://www.sciencedirect.com/science/article/pii/S0375960197001540}.

\bibitem{PhysRevLett.80.4361}
\bibinfo{author}{Habib, S.}, \bibinfo{author}{Shizume, K.} \&
  \bibinfo{author}{Zurek, W.~H.}
\newblock \bibinfo{title}{Decoherence, chaos, and the correspondence
  principle}.
\newblock \emph{\bibinfo{journal}{Phys. Rev. Lett.}}
  \textbf{\bibinfo{volume}{80}}, \bibinfo{pages}{4361--4365}
  (\bibinfo{year}{1998}).
\newblock \urlprefix\url{http://link.aps.org/doi/10.1103/PhysRevLett.80.4361}.

\bibitem{PhysRevLett.60.1836}
\bibinfo{author}{Wolinsky, M.} \& \bibinfo{author}{Carmichael, H.~J.}
\newblock \bibinfo{title}{Quantum noise in the parametric oscillator: From
  squeezed states to coherent-state superpositions}.
\newblock \emph{\bibinfo{journal}{Phys. Rev. Lett.}}
  \textbf{\bibinfo{volume}{60}}, \bibinfo{pages}{1836--1839}
  (\bibinfo{year}{1988}).
\newblock \urlprefix\url{http://link.aps.org/doi/10.1103/PhysRevLett.60.1836}.

\bibitem{PhysRevA.69.043804}
\bibinfo{author}{Everitt, M.~J.} \emph{et~al.}
\newblock \bibinfo{title}{Superconducting analogs of quantum optical phenomena:
  Macroscopic quantum superpositions and squeezing in a superconducting
  quantum-interference device ring}.
\newblock \emph{\bibinfo{journal}{Phys. Rev. A}} \textbf{\bibinfo{volume}{69}},
  \bibinfo{pages}{043804} (\bibinfo{year}{2004}).
\newblock \urlprefix\url{http://link.aps.org/doi/10.1103/PhysRevA.69.043804}.

\bibitem{PhysRevA.62.054101}
\bibinfo{author}{Nogues, G.} \emph{et~al.}
\newblock \bibinfo{title}{Measurement of a negative value for the wigner
  function of radiation}.
\newblock \emph{\bibinfo{journal}{Phys. Rev. A}} \textbf{\bibinfo{volume}{62}},
  \bibinfo{pages}{054101} (\bibinfo{year}{2000}).
\newblock \urlprefix\url{http://link.aps.org/doi/10.1103/PhysRevA.62.054101}.

\bibitem{Biaynicki-Birula2002}
\bibinfo{author}{Bia\l{}ynicki-Birula, I.}, \bibinfo{author}{Cirone, M.},
  \bibinfo{author}{Dahl, J.}, \bibinfo{author}{Fedorov, M.} \&
  \bibinfo{author}{Schleich, W.}
\newblock \bibinfo{title}{{In- and Outbound Spreading of a Free-Particle
  s-Wave}}.
\newblock \emph{\bibinfo{journal}{Physical Review Letters}}
  \textbf{\bibinfo{volume}{89}}, \bibinfo{pages}{060404}
  (\bibinfo{year}{2002}).
\newblock
  \urlprefix\url{http://link.aps.org/doi/10.1103/PhysRevLett.89.060404}.

\bibitem{PhysRevB.82.014512}
\bibinfo{author}{Kumar, S.} \& \bibinfo{author}{DiVincenzo, D.~P.}
\newblock \bibinfo{title}{Exploiting kerr cross nonlinearity in circuit quantum
  electrodynamics for nondemolition measurements}.
\newblock \emph{\bibinfo{journal}{Phys. Rev. B}} \textbf{\bibinfo{volume}{82}},
  \bibinfo{pages}{014512} (\bibinfo{year}{2010}).
\newblock \urlprefix\url{http://link.aps.org/doi/10.1103/PhysRevB.82.014512}.

\bibitem{PhysRevB.82.220505}
\bibinfo{author}{Deng, C.}, \bibinfo{author}{Gambetta, J.~M.} \&
  \bibinfo{author}{Lupa\ifmmode~\mbox{\c{s}}\else \c{s}\fi{}cu, A.}
\newblock \bibinfo{title}{Quantum nondemolition measurement of microwave
  photons using engineered quadratic interactions}.
\newblock \emph{\bibinfo{journal}{Phys. Rev. B}} \textbf{\bibinfo{volume}{82}},
  \bibinfo{pages}{220505} (\bibinfo{year}{2010}).
\newblock \urlprefix\url{http://link.aps.org/doi/10.1103/PhysRevB.82.220505}.

\bibitem{PhysRevB.74.224506}
\bibinfo{author}{Wallquist, M.}, \bibinfo{author}{Shumeiko, V.~S.} \&
  \bibinfo{author}{Wendin, G.}
\newblock \bibinfo{title}{Selective coupling of superconducting charge qubits
  mediated by a tunable stripline cavity}.
\newblock \emph{\bibinfo{journal}{Phys. Rev. B}} \textbf{\bibinfo{volume}{74}},
  \bibinfo{pages}{224506} (\bibinfo{year}{2006}).
\newblock \urlprefix\url{http://link.aps.org/doi/10.1103/PhysRevB.74.224506}.

\bibitem{PhysRevB.63.144530}
\bibinfo{author}{Everitt, M.~J.} \emph{et~al.}
\newblock \bibinfo{title}{Fully quantum-mechanical model of a squid ring
  coupled to an electromagnetic field}.
\newblock \emph{\bibinfo{journal}{Phys. Rev. B}} \textbf{\bibinfo{volume}{63}},
  \bibinfo{pages}{144530} (\bibinfo{year}{2001}).
\newblock \urlprefix\url{http://link.aps.org/doi/10.1103/PhysRevB.63.144530}.

\bibitem{PhysRevB.64.184517}
\bibinfo{author}{Everitt, M.~J.} \emph{et~al.}
\newblock \bibinfo{title}{Quantum statistics and entanglement of two
  electromagnetic field modes coupled via a mesoscopic squid ring}.
\newblock \emph{\bibinfo{journal}{Phys. Rev. B}} \textbf{\bibinfo{volume}{64}},
  \bibinfo{pages}{184517} (\bibinfo{year}{2001}).
\newblock \urlprefix\url{http://link.aps.org/doi/10.1103/PhysRevB.64.184517}.

\bibitem{PhysRevB.72.014508}
\bibinfo{author}{Stiffell, P.~B.}, \bibinfo{author}{Everitt, M.~J.},
  \bibinfo{author}{Clark, T.~D.}, \bibinfo{author}{Harland, C.~J.} \&
  \bibinfo{author}{Ralph, J.~F.}
\newblock \bibinfo{title}{Quantum downconversion and multipartite entanglement
  via a mesoscopic superconducting quantum interference device ring}.
\newblock \emph{\bibinfo{journal}{Phys. Rev. B}} \textbf{\bibinfo{volume}{72}},
  \bibinfo{pages}{014508} (\bibinfo{year}{2005}).
\newblock \urlprefix\url{http://link.aps.org/doi/10.1103/PhysRevB.72.014508}.

\bibitem{PhysRevA.48.2494}
\bibinfo{author}{Wielinga, B.} \& \bibinfo{author}{Milburn, G.~J.}
\newblock \bibinfo{title}{Quantum tunneling in a kerr medium with parametric
  pumping}.
\newblock \emph{\bibinfo{journal}{Phys. Rev. A}} \textbf{\bibinfo{volume}{48}},
  \bibinfo{pages}{2494--2496} (\bibinfo{year}{1993}).
\newblock \urlprefix\url{http://link.aps.org/doi/10.1103/PhysRevA.48.2494}.

\bibitem{PhysRevA.70.052105}
\bibinfo{author}{Santamore, D.~H.}, \bibinfo{author}{Goan, H.-S.},
  \bibinfo{author}{Milburn, G.~J.} \& \bibinfo{author}{Roukes, M.~L.}
\newblock \bibinfo{title}{Anharmonic effects on a phonon-number measurement of
  a quantum-mesoscopic-mechanical oscillator}.
\newblock \emph{\bibinfo{journal}{Phys. Rev. A}} \textbf{\bibinfo{volume}{70}},
  \bibinfo{pages}{052105} (\bibinfo{year}{2004}).
\newblock \urlprefix\url{http://link.aps.org/doi/10.1103/PhysRevA.70.052105}.

\bibitem{Bell}
\bibinfo{author}{Bell, J.}
\newblock \bibinfo{title}{Against `measurement'}.
\newblock \emph{\bibinfo{journal}{Physics World}} \textbf{\bibinfo{volume}{3}},
  \bibinfo{pages}{33} (\bibinfo{year}{1990}).

\bibitem{RevModPhys.76.1267}
\bibinfo{author}{Schlosshauer, M.}
\newblock \bibinfo{title}{Decoherence, the measurement problem, and
  interpretations of quantum mechanics}.
\newblock \emph{\bibinfo{journal}{Rev. Mod. Phys.}}
  \textbf{\bibinfo{volume}{76}}, \bibinfo{pages}{1267--1305}
  (\bibinfo{year}{2005}).
\newblock \urlprefix\url{http://link.aps.org/doi/10.1103/RevModPhys.76.1267}.

\bibitem{MJENJPQCT}
\bibinfo{author}{Everitt, M.~J.}
\newblock \bibinfo{title}{On the correspondence principle: implications from a
  study of the nonlinear dynamics of a macroscopic quantum device}.
\newblock \emph{\bibinfo{journal}{New Journal of Physics}}
  \textbf{\bibinfo{volume}{11}}, \bibinfo{pages}{013014}
  (\bibinfo{year}{2009}).
\newblock \urlprefix\url{http://stacks.iop.org/1367-2630/11/i=1/a=013014}.

\bibitem{PhysRevA.79.032328}
\bibinfo{author}{Everitt, M.~J.}, \bibinfo{author}{Munro, W.~J.} \&
  \bibinfo{author}{Spiller, T.~P.}
\newblock \bibinfo{title}{Quantum-classical crossover of a field mode}.
\newblock \emph{\bibinfo{journal}{Phys. Rev. A}} \textbf{\bibinfo{volume}{79}},
  \bibinfo{pages}{032328} (\bibinfo{year}{2009}).
\newblock \urlprefix\url{http://link.aps.org/doi/10.1103/PhysRevA.79.032328}.

\end{thebibliography}

\end{document}

\title{ {\em Cool for cats} Supplementary Information:\\
Engineering two photon decay in circuit QED. }

\author{M. J.  Everitt$^1$, T.P. Spiller$^2$, G . J. Milburn$^3$, R.D. Wilson$^1$ and A.M. Zagoskin}

\affiliation{$^1$Department of Physics, Loughborough University,
Loughborough, Leics LE11 3TU, United Kingdom,\\
$^2$Quantum Information Science, School of Physics and Astronomy,
University of Leeds, Leeds LS2 9JT, United Kingdom\\
$^3$Centre for engineered Quantum Systems, School of Mathematics and Physics, The University of Queensland, St Lucia, QLD 4072, Australia.
}

\begin{abstract}
We show that a microwave superconducting cavity can be engineered to have a dominant two photon decay term using two
cavities coupled by a SQUID. 
\end{abstract}

\maketitle

There are a number of models\cite{divincenzo,gambetta} whereby two microwave superconducting cavities can be nonlinerally coupled using SQUIDs. We will base our 
discussion on Kumar and Divincenzo\cite{divincenzo}. In that model, the hamiltonian describing two microwave cavities, a probe (p)  cavity and a signal (s) cavity, 
coupled with a SQUID is 
\begin{eqnarray}
H & = & E_{Cp}n_{p}^2+E_{Lp}\phi_p^2+E_{Cs}n_s^2+E_{Ls}\phi_s^2 \nonumber \\ \nonumber
& & \mbox{}+A\left [E^4_{Lp}\phi^4_{p}\cos^4\beta +E^4_{Ls}\phi^4_{s} \right.\sin^4\beta \\
&& +\left. 6E^2_{Lp}E^2_{Ls}\phi^2_{p}\phi^2_{s}\cos^2\beta\sin^2\beta\right ] \
\end{eqnarray} 
where $n_\alpha,\phi_\alpha$ are the standard charge and phase conjugate variables describing the collective electrical degree of freedom  in each cavity and $A=16\pi^2L_1/\Phi_0^4$ with $L_1$
defined as the coefficient of the leading non-linear current term of the SQUID inductance. 
We will set $\cos^2\beta =\sin^2\beta =1/2$. 

The system can be quantised in the 
usual way in terms of the bosonic annihilation and creation operators $b,b^\dagger$ for the probe and $a,a^\dagger$ and for the signal cavity defined by\cite{Wallquist}
\begin{eqnarray}
\phi_p & \rightarrow &   \left (\frac{E_{Cp}}{4E_{Lp}}\right )^{1/4}(b+b^\dagger)\\
n_p & \rightarrow &  -i \left (\frac{E_{Lp}}{4E_{Cp}}\right )^{1/4}(b-b^\dagger)\\
\phi_s & \rightarrow &   \left (\frac{E_{Cs}}{4E_{Ls}}\right )^{1/4}(a+a^\dagger)\\
n_s & \rightarrow &   -i\left (\frac{E_{Ls}}{4E_{Cs}}\right )^{1/4}(a-a^\dagger)
\end{eqnarray}
The Hamiltonian may then be written as 
\begin{eqnarray}
H & = & \hbar\omega_pb^\dagger b+\hbar \omega_s a^\dagger a+\hbar\chi_b b^{\dagger\ 2}b^2 +\hbar\chi_a a^{\dagger\ 2}a^2 \nonumber\\
& & \mbox{}+\hbar\sqrt{\chi_a\chi_b}\left (b^2a^{\dagger \ 2}+b^{\dagger\ 2}a^2+4a^\dagger ab^\dagger b\right )
\end{eqnarray}
Unlike Kumar and Divincenzo\cite{divincenzo} we have {\em not} neglected the terms like $b^2a^{\dagger \ 2}$ as we will choose
$\omega_p=\omega_s$ so that these terms are resonant. 

We now include the dissipative channels for this model in the usual way. The density operator for the total system, in the interaction picture, satisfies
\begin{equation}
\frac{d\rho}{dt} =-i[H_I,\rho]+\kappa_a{\cal D}[a]\rho+\kappa_b{\cal D}[b]\rho
\end{equation}
where ${\cal D}[L]\rho=L\rho L^{\dag}-\frac{1}{2} (L^\dag L \rho + \rho L^{\dag}L)$ and 
\begin{multline}
\label{int-ham}
H_I = \hbar \chi_b b^{\dagger\ 2}b^2 + \hbar \chi_a a^{\dagger\ 2}a^2 +\hbar(\epsilon^*b+\epsilon b^\dagger) \\+ \hbar \sqrt{\chi_a\chi_b}\left (b^2a^{\dagger \ 2}+b^{\dagger\ 2}a^2+4a^\dagger ab^\dagger b\right )
\end{multline}
and $\kappa_a,\kappa_b$ are the decay rates of the photon number in the signal and probe cavity respectively and we have included a resonant coherent driving of the probe cavity with  $\epsilon=\sqrt{\kappa_b}{\varepsilon_b}$ where $|\varepsilon_b|^2$ is the photon flux of the driving field. 
We have also assumed that each cavity sees a zero temperature environment. 

In the absence of the SQUID mediated interactions the probe cavity will relax to a coherent state with the steady state amplitude 
\begin{equation}
\beta_0=\frac{-2i\epsilon}{\kappa_b}
\end{equation}
We will chose the phase of the probe driving as a reference phase and set $\beta_0$ to be real. 
If we make a canonical transformation to the displaced picture by
\begin{equation}
b=\bar{b}+\beta_0
\end{equation}
we can linearise the Hamiltonian, Eq. \ref{int-ham},  in  $\bar{b},\bar{b}^\dagger$ to  obtain
\begin{equation}
\label{linearised}
H_I=H_a+4\hbar \sqrt{\chi_a\chi_b}\beta_0(\bar{b}+\bar{b}^\dagger)a^\dagger a+2\hbar\sqrt{\chi_a\chi_b}\beta_0(\bar{b}^\dagger a^2+\bar{b} a^{\dagger \ 2})
\end{equation}
where the effective Hamiltonian for the signal mode alone is 
\begin{equation}
H_a =\hbar\chi_a a^{\dagger\ 2}a^2+4\hbar\sqrt{\chi_a\chi_b}\beta_0^2 a^\dagger a +\hbar\sqrt{\chi_a\chi_b}\beta_0^2(a^2+a^{\dagger\ 2})
\end{equation}
which is equivalently to a parametrically driven Kerr non-linear cavity. This model was considered by Wielinga and Milburn\cite{wielinger}. It is equivalent to a double well system with a hyperbolic fixed point at the origin in phase space and two elliptic fixed points symmetrically displaced from the origin.  The second and third terms in Eq. (\ref{linearised}) can be given a familiar interpretation. The second term is of the same form as the radiation pressure interaction between a mechanical resonator ($\bar{b},\bar{b}^\dagger$) and a cavity field $(a,a^\dagger$). The last  term is equivalent to the quantum derivation of sub/second harmonic generation considered by Drummond et al.\cite{DMW2}.

\begin{figure}[!tb]
\begin{center}
\includegraphics[width=0.45\textwidth]{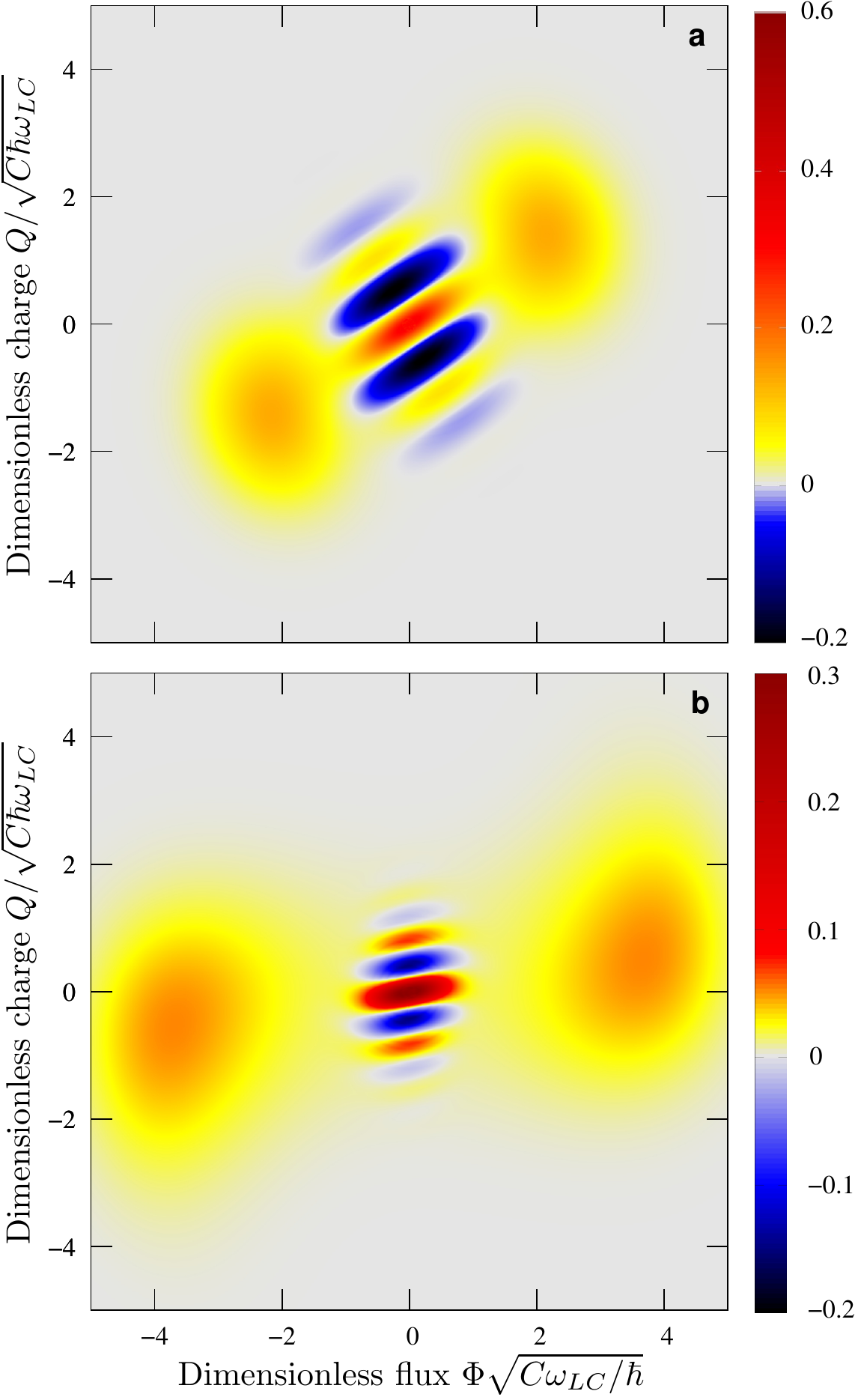}
\caption{\textbf{A persistent cat} Here we look at the effect of a including the dephasing term in addition to the  the bath of two-photon absorbers to the ring initially in its ground energy eigenstate. \textbf{\textsf{a}} The steady state solution for the environment as derived in this supplementary material and described by master equation Eq.~(\ref{mastereq}) and (\ref{ratios}) with $\Gamma_{2}=0.2$  (or $L_2=\sqrt{0.2}a^2$) and $\Gamma_{\perp}=0.05$ (or $L_\perp=\sqrt{0.05}a^\dag a$) and we have set the damping coefficient $\kappa_{a}=0$. \textbf{\textsf{b}} We  show that even for an environment, other than the one considered in \textsf{a}, where  dephasing  dominates over the the two-photon absorption process ($L_2=\sqrt{0.02}a^2$ and $L_\perp=\sqrt{0.08}a^\dag a$) it is still possible for the steady state of the ring to be a cat. }
\label{Fig:PurrCat}   
\end{center}
\end{figure}
We now assume that $\kappa_b$, the line width of the probe cavity is large, $\kappa_b >>\kappa_a,\sqrt{\chi_a\chi_b}$ and we adiabatically eliminate it from the dynamics.  In that case from the point of view of the signal mode,  the first term in Eq.(\ref{linearised}) looks like a fluctuating cavity detuning while the last arms looks like a two photon loss term.  This can be verified by explicit adiabatic elimination of the probe cavity field.  We assume that the probe cavity, in the displaced picture, remains very close to its steady state of zero photons. The method is described in \cite{Santamore}. The effective master equation for the signal cavity is 
\begin{equation}\label{mastereq}
\frac{d\rho_s}{dt} = -\frac{i}{\hbar}[H_a,\rho_s]+\Gamma_2{\cal D}[a^2]\rho_s+\Gamma_\perp{\cal D}[a^\dagger a]\rho_s+\kappa_a{\cal D}[a]\rho_s
\end{equation}
where the two photon decay rate $\Gamma_2$ and dephasing rate $\Gamma_\perp$ are given by
\begin{equation}\label{ratios}
\begin{split}
\Gamma_2 & =  \frac{16\chi_a\chi_b\beta_0^2}{\kappa_b} \\
\Gamma_\perp & =  \Gamma_2/4 =\frac{4\chi_a\chi_b\beta_0^2}{\kappa_b}
\end{split}
\end{equation}

A peculiar feature of using SQUID coupled cavities is that the  price paid for two-photon decay is an additional dephasing term on the signal cavity field.   Using the strong dependance on the steady state amplitude $\beta_0$ in the two photon rate  we can make this term dominate over the single photon decay of the signal cavity over the time scales of interest.  In Fig.~\ref{Fig:PurrCat} we show that the dephasing term that is introduced in the above (un-damped, $\kappa_{a}=0$) master equation has little effect on the  Schr\"odinger cat nature of the steady state solution associated with the two-photon absorbing bath. We therefore believe that the discussion in the main article is  in-line with the behaviour of realistic environments.